\documentclass[5p,twocolumn]{elsarticle}

\usepackage{hyperref}

\newcommand{\hl}[1]{#1}
\newcommand{\mathcolorbox}[2]{#2}
\newcommand{\colorbox}[2]{#2}
\newcommand{\fcolorbox}[3]{#3}

\journal{Chemical Engineering Science}

\usepackage{amsmath} 			
\usepackage{amssymb} 			
\usepackage{cases} 				
\usepackage{siunitx}			
\newcommand*{\pd}[2]{\frac{\partial\, #1}{\partial\, #2}}
\newcommand*{\fcn}[2]{#1 \big(\,#2\,\big)}
\newcommand*{\smallfcn}[2]{#1\! \left(#2\right)}
\newcommand*{\Fcn}[2]{#1 \Big(\,#2\,\Big)}
\newcommand*{\LM}[2]{\fcn{\mathcal{LM}}{#1,\,#2}}
\newcommand*{\GM}[2]{\fcn{\mathcal{GM}}{#1,\,#2}}
\newcommand*{\AM}[2]{\fcn{\mathcal{AM}}{#1,\,#2}}
\newcommand*{\WM}[2]{\fcn{\mathcal{WM}}{#1,\,#2,\, \beta}}
\newcommand*{\Qd}{\dot{Q}}
\newcommand*{\Hd}{\dot{H}}
\newcommand*{\md}{\dot{m}}
\newcommand*{\res}[1]{\mathcal{R}_{#1}}
\newcommand*{\Td}[1]{\dot{T}_{#1}}
\newcommand*{\DT}[1]{\Delta T_{#1}}
\newcommand*{\T}[1]{T_{#1}}
\newcommand*{\ew}[1]{e_{w #1}}
\newcommand*{\aAc}{\left(\alpha A\right)_{m,c}}
\newcommand*{\aAh}{\left(\alpha A\right)_{m,h}}
\newcommand*{\aA}{\left(\alpha A\right)_{m}}
\newcommand*{\kA}{\left(k A\right)_m}

\newcommand*{\cph}{\theta_3}
\newcommand*{\cpc}{\theta_4}
\newcommand*{\cphs}{\theta_5}
\newcommand*{\cpcs}{\theta_6}
\newcommand*{\mcpc}{\md_c\cdot \cpc}
\newcommand*{\mcph}{\md_h\cdot \cph}

\newcommand*{\mcpcsND}{\md_c \cpcs}
\newcommand*{\mcphsND}{\md_h \cphs}

\newcommand*{\Htilde}{\stackrel{\sim}{\smash{\mathcal{H}}\rule{0pt}{1.1ex}}}
\newcommand*{\Qtilde}{\stackrel{\sim}{\smash{\mathcal{Q}}\rule{0pt}{1.1ex}}}
\newcommand*{\Rtilde}{\stackrel{\sim}{\smash{\mathcal{R}}\rule{0pt}{1.1ex}}}
\renewcommand*{\vec}[1]{\underline{#1}}
\newcommand*{\mat}[1]{\mathbf{#1}}
\newcommand*{\p}{\vec{\upsilon}}
\newcounter{saveenum}

\begin{document}

\begin{frontmatter}

\title{A Low-Order Dynamic Model of Counterflow Heat Exchangers for the Purpose of Monitoring Transient and Steady-State Operating Phases}

\author{Maik Gentsch\corref{cor1}%
	\fnref{fn1}}
\ead{maik.gentsch@tu-berlin.de}

\author{Rudibert King\fnref{fn2}}
\ead{rudibert.king@tu-berlin.de}

\cortext[cor1]{Corresponding author}
\fntext[fn1]{Graduate Research Assistant}
\fntext[fn2]{Head of Department \\ \textcopyright\ 2020. This manuscript version is made available under the CC BY-NC-ND 4.0 license \url{https://creativecommons.org/licenses/by-nc-nd/4.0/}\ .}

\address{Technische Universit\"at Berlin, Chair of Measurement and Control, Straße des 17. Juni 135, 10623 Berlin, Germany}

%
%

\begin{abstract}
	\hl{We present a model-based real-time method to monitor a
	counterflow heat exchanger's thermal performance for all operating
	conditions.
	A first principle reference model that describes the reference
	counterflow process in an accurate manner is derived first.
	Real gas behavior is taken into account.
	Without simplifications, the respective equations must be solved in an
	iterative, computationally expensive manner, which prohibits their use
	for real-time monitoring purposes.
	Therefore, we propose one-step-solvable model equations, resulting in an
	approximate but quick model, which is able to track an important thermal
	property reliably.
	The monitoring, i.e., the online estimation of the thermal properties, is
	achieved via a nonlinear Kalman-Filter.
	Due to the low-order dynamic model formulation, the overall monitoring
	scheme is accompanied by an acceptable computational burden.
	Moreover, it is easy to deploy and to adapt in industrial practice.
	Monitoring results, where the reference model replaces a real process with supercritical carbon dioxide, are given and discussed herein.
}

\end{abstract}

\begin{keyword}
Heat exchanger modeling \sep
Model-based supervision \sep
Flexible operation \sep
Real gas process fluid \sep
Online monitoring \sep
Extended Kalman Filter
\end{keyword}

\end{frontmatter}


\section{Introduction}
\label{sec:intro}

\hl{Due to modern standards and future challenges, industrial plants will be bound to run very flexibly across a wide range of operating points without any trade-offs concerning their availability and reliability.
This is true, for example, for multi-stage compressors with intermediate heat exchanger units.}
Formerly, these machines were designed for steady-state operating conditions.
This, likewise, applies to the supervision methodology of such machines in industrial practice.
Typically, supervision is based on measurements only, alarming the supervisor if data exceed certain thresholds concerning the expected steady-state operating values.
Using this methodology during transient operating phases could lead to frequent false alarms if the supervised machine comprises any dynamics within the relevant time scale.

In this paper, we investigate \hl{model-based supervision of} counterflow heat exchangers that are known to be a sluggish plant component when it comes to industrial scale.
\hl{Therefore, a proper mathematical model description is essential.
Since transient behavior is to be covered and online monitoring is addressed, the model must consider relevant system dynamics and, further, the computational effort to solve the model equations should be low at the same time.}

The modeling of heat exchangers has been proposed in various studies, from simple lumped descriptions \cite{yin2003analytic,laszczyk2017simplified} to complex spatial models using Computational Fluid Dynamics \cite{bhutta2012cfd}.
The most common application considers one-dimensional parameter and temperature distributions represented by partial differential equations, which are solved by implementing a discretization scheme, i.e., finite volume or finite difference methods \cite{kaern2011experimental,alobaid2017progress,correa1987dynamic,coletti2011aDynamic}.
Usually, model building and equation solving is performed in the framework of technically mature software, such as Dymola (e.g., \cite{kaern2011experimental,sodja2009some})\hl{ or gPROMS (e.g., \mbox{\cite{coletti2011aDynamic}}).} 
Because of the heavy impact on plant efficiency, further investigations have been conducted to model the phenomenon of fouling in more detail (e.g., \cite{ishiyama2010impact,coletti2011aDynamic}).
The major purposes of the more complex models stated above are: i) achieving deeper system knowledge; ii) accomplishing off-line analyses; and iii) improving design methods for heat exchangers.
Unfortunately, they are accompanied by a tremendous effort concerning model implementation and parametrization, resulting in a highly customized application for a specific heat exchanger.

\hl{The aim of this work is to meet certain industrial requirements, namely, adaptability and ease of deployment. 
More specifically, these requirements include: i) little numerical effort; ii) an algorithm proven to be reliable in terms of numerical issues; and iii) the simplicity of adaption to various heat exchangers.
Though the simpler, low-order models found in literature may pass these criteria, they require some inappropriate assumptions concerning the scope of the application addressed here, e.g., the neglect of specific temperature differences between the heat exchanger's intake and outlet cross-section.
Therefore, we decided against a model based on partial differential equations, and derived a suitable low-order dynamic model based on simple first principles with few global parameters.
However, as the model does not consider spatial temperature distributions, some heuristics are utilized to form the dynamic model equations.
The model formulation allows for incorporating arbitrary enthalpy calculation models, which is a further feature of this work.
On that account, we were able to present the effect of the common perfect gas assumption on the monitoring results when applied to real gas applications, such as the supercritical carbon dioxide heat exchanger from the simulation study below.
}

Beside fouling, there are conceivable faults like coolant leakage or faults of the neighboring plant components (e.g., compressor, valves) that could lead to a significant change of the heat exchanger's performance.
Fault detection and isolation of a specific fault, however, is not within the scope of this paper.
Instead, \hl{the proposed monitoring scheme provides the time series of parameters, filtered measurements, and, consequently, residuals between real and filtered measurements.
All of them could serve as a base for such fault detection and isolation algorithms.
The provision of these time series in real-time, i.e., during plant operation, necessitates proper estimation techniques.
For this purpose, the Extended Kalman Filter scheme is applied.
This scheme considers the model equations and the respective algorithm is of moderate numerical complexity, as is it formulated in a recursive manner.
}


\hl{Within the paper the following issues are presented and discussed:}
In Section \ref{sec:scope}, the scope of the application is stated, and all assumptions used are summarized and discussed.
\hl{The model building is presented }in Section \ref{sec:model}.
Because the final aim is to monitor unmeasurable properties, i.e., (convective) heat transfer coefficients, we develop a reference model for validation purposes in Section \ref{sec:model} first.
The reference model is accurate if the reference counterflow process, as defined in Section \ref{sec:scope}, is valid.
We derive the respective model equations and give a proof of stability, which is an essential property for the derivation of the low-order state equations.
\hl{Based on the reference model, an approximate model is derived, which is suitable for the real-time application of online monitoring due to its significantly lower computational complexity.}
In this paper, the term ``monitoring'' is the equivalent of an online parameter estimation on the basis of a model-based measurement scheme consisting of the dynamic model as a part of an Extended Kalman Filter.
The whole approach is presented in Section \ref{sec:monitoring}.
The main results of this investigation will be discussed in the last subsections before the conclusions are drawn in Section \ref{sec:conclusion}.

\section{Assumptions and the Scope of the Application}
\label{sec:scope}

\subsection{Reference Counterflow Process}
\label{sec:ReferenceProcess}

\begin{figure}
	\begin{center}
		\includegraphics{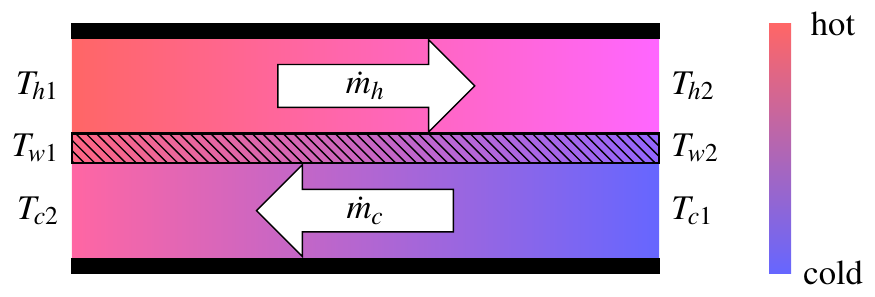}
	\end{center}
	\caption[Reference counterflow process]{Reference counterflow process: \\  h -- hot fluid, c -- cold fluid, w -- wall, 1 -- intake, 2 -- outlet}
	\label{fig:counterflow}
\end{figure}

The presented modeling approach addresses all types of two-fluid heat exchangers fitting the reference counterflow process, as shown in Fig.\ \ref{fig:counterflow}.
The model allows for lumped temperature information on the heat exchanger's intake and outlet cross-section; hence, there is neither need nor chance to interact with spatial parameter or temperature distributions.
Because no further assumption concerning the inner geometry of the exchanger is made, one could fit shell-and-tube heat exchangers into the illustrated scheme by merging the inner tube bundle to a single wall and, accordingly, the inner tube streams to a single flow.
This appears to be valid as long as 
\begin{enumerate}
	\item the process fluid enters the exchanger near the coolant exit and vice versa,
	\label{item:counterflow}
	\setcounter{saveenum}{\value{enumi}}
\end{enumerate}
which is quite a relaxation compared to a strict counterflow definition.
Despite the lumped temperature approach, we do not think of the inner wall as a lumped mass with a single temperature value.
In fact, a wall temperature distinction is drawn between the intake and outlet cross-section, which is in accordance with the expected situation drafted in Fig.\ \ref{fig:counterflow}.
This truly differs from the conventional lumped models.
On the other hand, a local wall temperature distribution in a radial direction is neglected, which means
\begin{enumerate}
	\setcounter{enumi}{\value{saveenum}}
	\item the local thermal conductance of the inner wall in a radial direction is infinite.
	\label{item:InfWallCond}
	\setcounter{saveenum}{\value{enumi}}
\end{enumerate}
Further assumptions defining the reference process include:
\begin{enumerate}
	\setcounter{enumi}{\value{saveenum}}
	\item the outer casing is adiabatic; 	\label{item:adiabat}
	\item the interchange of heat is an isobaric process; and 	\label{item:isobar}
	\item the response time, due to the thermal inertia of the inner wall, is many times greater than the delay due to mass transport.	\label{item:NoDelay}
	\setcounter{saveenum}{\value{enumi}}
\end{enumerate}
For supervision, two different rating problems could be thought of: the determination of heat transfer (thermal rating) and the pressure drop performance (hydraulic rating) \cite{shah2003fundamentals}.
In this research, we focus on the carbon dioxide process fluid at a supercritical state, which appears, for example, in a carbon capture and storage (CCS) process with the aim of reducing greenhouse gases in the atmosphere.
Concerning the relative high pressure level, we found that the pressure drop over the exchanger was nearly negligible, which lead to assumption \ref{item:isobar} and the decision to accomplish a thermal rating solely.
However, in the presence of relevant pressure losses, we argue that a thermal rating is capable of reflecting a change of system behavior even if the underlaying model assumes isobaric conditions.
Note that we do not assume ideal gas behavior.

In terms of instrumentation, it is presupposed that
\begin{enumerate}
	\setcounter{enumi}{\value{saveenum}}
	\item mass flow, intake, and outlet temperatures, both of the process and cooling fluid, are known; and 	\label{item:Inputs}
	\item if one or both fluids shall be treated as a real gas, their respective pressure is known. 	\label{item:PreasMeas}
	\setcounter{saveenum}{\value{enumi}}
\end{enumerate}
Quite clearly, assumption \ref{item:Inputs} provides an ideal situation, which is hard to find in industrial practice.
Moreover, even if all temperature measurements exist, they are quite often delayed and considerably biased due to the thick-walled shield casings.
There are two ways to deal with missing or unreliable measurements when it comes to model validation:
replacement by data sheet specifications or replacement by peripheral model calculations.
Both of them bring some uncertainty, but we chose the latter approach.
Thus, the validation of the heat exchanger model is correlated with these peripheral models.
Because they are beyond the scope of this paper, we will present the whole issue of validation given an imperfect data status in a future article.
For the sake of deriving the model equations, we accept assumption \ref{item:Inputs}, stating that the model validation yields satisfactory results.
Within the monitoring scheme, that assumption could be relaxed since uncertain information is permitted.
Further details are provided in Section \ref{sec:reduced_information}.

\subsection{Thermal Rating}

The overall heat transfer coefficient $k_m$ serves as an appropriate measure to accomplish the thermal performance rating.
With a lack of information about the total heat transfer surface area $A$, it is more convenient to look for the overall thermal conductance $\kA$ as a single term:
\begin{equation}
\kA = \left| \frac{\Qd}{\DT{m}} \right| \, .
\label{eq:kA_stat}
\end{equation}
For heat exchangers, the mean temperature difference $\DT{m}$ is typically calculated using the logarithmic mean ($\mathcal{LM}$) of the fluid's temperature differences at the intake and outlet cross-section, respectively (cf.\ Fig.\ \ref{fig:counterflow}):
 \begin{align}
\DT{1} &= \T{h1} - \T{c2} \,, \qquad \DT{2} = \T{h2} - \T{c1} \, , 
\label{eq:DT12} \\
\DT{m} &= \LM{\DT{1}}{\DT{2}} := \frac{\DT{1}-\DT{2}}{\ln{\frac{\DT{1}}{\DT{2}}}} \, .
\label{eq:LM}
\end{align}
Even though the derivation of the log-mean temperature difference assumes strict counterflow or parallel flow conditions \cite{baehr1996warme}, it is commonly used for alternative flow arrangements, considering a hypothetical counterflow unit operating at the same resistance and effectiveness \cite{shah2003fundamentals}.

The aforementioned temperatures are accessible in all operating phases.
This does not apply to the total heat transfer rate $\Qd$, the calculation of which assumes a steady-state operation.
To overcome this issue, we look at the serial connection of thermal resistances, leading to:\footnote{
	Note that the thermal resistance of the heat transmitting wall is neglected due to assumption \ref{item:InfWallCond}.}
 \begin{align}
\kA = \left( \frac{1}{\aAh} + \frac{1}{\aAc} \right) ^{-1} \, ,
\label{eq:kA_dyn}
\end{align}
where $\aA$ represents the overall convection conductance at the hot fluid side (subscript $h$) and the cold fluid side (subscript $c$), respectively.
Their definitions do not depend on steady-state conditions:
 \begin{align}
\aAh &= \left| \frac{\Qd_h}{\LM{\DT{h1}}{\DT{h2}}} \right|  \, , 
\label{eq:aAh} \\
\aAc &= \left| \frac{\Qd_c}{\LM{\DT{c1}}{\DT{c2}}} \right|\, .
\label{eq:aAc}
\end{align}
The temperature differences as well as the heat transfer rates are no longer noted in reference to the opposite fluid stream but to the inner wall, leading to the introduction of wall temperatures at the exchanger's intake and outlet cross-section, as depicted in Fig.\ \ref{fig:counterflow}:
 \begin{align}
\DT{h1} &= \T{h1} - \T{w1} \, , & 	\DT{h2} &= \T{h2} - \T{w2} \, , 
\label{eq:DTh12} \\
\DT{c1} &= \T{w1} - \T{c2} \, , &	\DT{c2} &= \T{w2} - \T{c1} \, .
\label{eq:DTc12}
\end{align}
Commonly, wall temperature measurements do not exist; hence, we have derived a model to calculate these time-variant variables.
Note that additional terms for fouling resistances are not introduced.
As a matter of fact, a separate monitoring of $\aAh$ and $\aAc$ could provide more information than a coupled monitoring of $\kA$, according to Eq.\ (\ref{eq:kA_dyn}).
Unfortunately, the presented approach does not allow for decoupled monitoring under most operating conditions, as will be shown in Section \ref{sec:steady_transient_monitoring}.
However, the coupled estimate of $\kA$ is capable of tracking the exchanger's overall thermal performance quickly and reliably.

\section{Modeling}
\label{sec:model}

The starting point for a model-based monitoring approach is a nonlinear, dynamic system description
\begin{numcases}{}
\smallfcn{\vec{\dot{x}}}{t} = \fcn{\vec{f}}{\vec{x},\,\vec{u},\,\vec{\theta},\, t}\ , 	 & $\smallfcn{\vec{x}}{t_0}=\vec{x}_0$ , \label{eq:ZDGL} \\[.5em]
\smallfcn{\vec{y}}{t} = \fcn{\vec{g}}{\vec{x},\,\vec{u},\,\vec{\theta},\, t}\, ,  & \label{eq:Ausgangsgleichung}
\end{numcases}
where $\vec{y} \in \mathbb{R}^{n_y}$, $\vec{x} \in \mathbb{R}^{n_x}$, $\vec{u} \in \mathbb{R}^{n_u}$, and $\vec{\theta} \in \mathbb{R}^{n_\theta}$ are the measurable outputs, the states, the inputs, and the parameters of the model, respectively.
In general, all of these values are time-variant, but the model parameters are assumed to vary much slower than the other variables.
To increase the readability of the equations, the time argument $t$ is suppressed in what follows.

Only two dynamic variables, $\T{w1}$ and $\T{w2}$, composed in the state $\vec{x}$, are considered for an appropriate description to facilitate the setup of a real-time algorithm, namely, the wall temperatures used in Eqs.\ (\ref{eq:DTh12})--(\ref{eq:DTc12}).
In summary, the proposed assignment of variables for the system (\ref{eq:ZDGL})--(\ref{eq:Ausgangsgleichung}) is:
\begin{equation}
\vec{x} = 
\begin{bmatrix}
\T{w1} \\ \T{w2}
\end{bmatrix}\, , \quad
\vec{y} = 
\begin{bmatrix}
\T{h2} \\ \T{c2}
\end{bmatrix}\, , \quad
\vec{u} = 
\begin{bmatrix}
\T{h1} \\ \T{c1} \\ \md_h \\ \md_c
\end{bmatrix}\, , \quad
\vec{\theta} =
\begin{bmatrix}
\theta_1 \\ \theta_2 \\ \vdots \\ \theta_7
\end{bmatrix}\, .
\label{eq:y_x_u}
\end{equation}
Based on the state $\vec{x}$, the given inputs $\vec{u}$ (see assumption \ref{item:Inputs}), and parameters $\vec{\theta}$, the outlet temperatures $\vec{y}$ are calculated with the output equation $\vec{g}$.
The supervision scheme will utilize the residual between these model outputs and measured outlet temperatures to calculate a proper estimate of the model parameters used for the thermal performance rating.
In the input vector, $\md_h$ and $\md_c$ are the hot and cold fluids' mass flows, respectively.
Like the model equations $\vec{g}$ and $\vec{f}$, the concrete model parameters $\vec{\theta}$ differ depending on whether they belong to the reference or the approximate model.
They will be introduced later.

In Section \ref{sec:scope}, it was pointed out that the sluggish response due to the thermal inertia of the inner wall is significantly more dominant than secondary dynamics, like the mass transport delay (see assumption \ref{item:NoDelay}).
To that end, we use a quasi-steady-state approach.
That means the outlet temperatures $\vec{y}$ are calculated as if they would adjust after an infinite amount of time but under the conditions of fixed wall temperatures $\vec{x}$ and fixed inputs $\vec{u}$.
As long as this calculation leads to an imbalance between the wall heating and cooling fluxes, the wall temperatures will tend toward their respective equilibrium state, affecting the subsequent quasi-steady-state calculation of $\vec{y}$.

We presuppose the existence of an appropriate model to calculate specific enthalpies of the process and the cooling fluids.
For the sake of generality, the dependancy on pressure is considered, and the enthalpy calculation is denoted as
\begin{equation}
\fcn{h_h}{T,\, p} \quad \text{and} \quad \fcn{h_c}{T,\, p}\, .
\end{equation}
With respect to availability and accuracy, one has to decide whether to use a calorically perfect gas, an incompressible fluid, a thermally perfect gas, or a real gas model.
Deviations from the real fluid behavior will affect the quantity of the observed thermal property, as will be shown in Section \ref{sec:steady_transient_monitoring}.
Before the dynamic model $\vec{f}$ is specified in Section \ref{sec:ZDGL}, the output equations $\vec{g}$ and the steady-state solutions for the reference and approximate models are introduced each in Sections \ref{sec:OutputEq} and \ref{sec:steady}, respectively.

\subsection{Output Equations}
\label{sec:OutputEq}

\subsubsection{Reference Model}
\label{sec:OutputEq_reference}

As mentioned in \cite{shah2003fundamentals}, only two important relationships constitute the entire thermal design procedure (or, vice versa, the thermal rating problem) of two-fluid heat exchangers.
The first of them is the heat transfer rate equation represented by Eq.\ (\ref{eq:kA_stat}) for both fluids (steady-state only) or by Eqs.\ (\ref{eq:aAh})--(\ref{eq:aAc}) in a partitioned manner.
In a highly transient situation, with fast changing intake temperatures, we have to consider cases $\DT{h1}<0$ or $\DT{c2}<0$, for which the logarithmic mean according to Eq.\ (\ref{eq:LM}) is not defined due to arguments with opposite signs.
To be able to describe such phases as well, we use an unrestricted formulation:\footnote{
	We always balance from the fluid point of view. Thus, a negative $\Qd$ or $\Hd$ denotes a cooling of the fluid.}
 \begin{align}
&\Qd_h = -\fcn{\mathcal{Q}}{\DT{h1},\,\DT{h2},\,\aAh} \, , \label{eq:Qd_extended}\\
&\Qd_c = \phantom{- } \fcn{\mathcal{Q}}{\DT{c1},\,\DT{c2},\,\aAc} \, , \label{eq:Qc_extended} \\[.5em]
&\fcn{\mathcal{Q}}{z_1,\,z_2,\,z_3} :=
\begin{cases}
z_3 \cdot \LM{z_1}{z_2} & ,\ \left( z_1,\, z_2 \right) \in \mathbf L_{z_1}^{z_2} \\
z_3 \cdot \AM{z_1}{z_2} & ,\ \left( z_1,\, z_2 \right) \notin \mathbf L_{z_1}^{z_2} 
\end{cases} \, ,\label{eq:Qd_unrestricted}
\end{align}
where
 \begin{align}
&\AM{z_1}{z_2} := \frac{z_1+z_2}{2} \qquad \text{and}\\
&\mathbf L_{z_1}^{z_2} := \{ \left(z_1,\, z_2\right) \in \mathbb{R}^2 \ \vert \ z_1>0\, ,\ z_2>0\, ,\ z_1\ne z_2  \}
\end{align}
denote the arithmetic mean and the domain of the logarithmic mean, respectively.
Calligraphic variables, such as $\mathcal{Q}(z_1,z_2,z_3)$, denote functions that are specifically defined in this contribution to ensure a compact representation of relevant dependencies.

The second elementary relationship is given by the isobaric enthalpy rate equations
 \begin{align}
\Hd_h &= \fcn{\mathcal{H}_h}{\T{h2}}:=\md_h\cdot \left[ \fcn{h_h}{\T{h2},\, p_h} - \fcn{h_h}{\T{h1},\, p_h} \right]\, , \label{eq:Hhdot}\\
\Hd_c &= \fcn{\mathcal{H}_c}{\T{c2}}:=\md_c\cdot \left[ \fcn{h_c}{\T{c2},\, p_c} - \fcn{h_c}{\T{c1},\, p_c} \right]\, . \label{eq:Hcdot}
\end{align}
Because of the adiabatic outer casing, the heat transfer rates and enthalpy rates must be of equal value.
Such equalities, which will appear in different forms in this contribution, can always be reformulated as a root searching task of an appropriate residual function by bringing all terms of an equality on one side.
For the specific case considered here, given the current wall temperatures $\vec{x}$, the intake temperatures, and the mass flows, all compressed in $\vec{u}$, in addition to the overall convection conductances
\begin{equation}
\theta_1 = \aAh \quad \text{and} \quad \theta_2 = \aAc\, ,
\end{equation}
which are treated as model parameters here, the roots of the residual functions
 \begin{align}
\fcn{\res{h}}{T_{h2}^*} &:= \fcn{\mathcal{H}_h}{\T{h2}^*} + \fcn{\mathcal{Q}}{\T{h1}-\T{w1},\,\T{h2}^*-\T{w2},\,\theta_1}\, ,
\label{eq:resh} \\
\fcn{\res{c}}{T_{c2}^*} &:= \fcn{\mathcal{H}_c}{\T{c2}^*} - \fcn{\mathcal{Q}}{\T{w1}-\T{c2}^*,\,\T{w2}-\T{c1},\,\theta_2}\, ,
\label{eq:resc}
\end{align}
have to be determined, which are the unknown outlet temperatures $\T{h2}$ and $\T{c2}$.
Note that with all remaining values fixed, $\res{h}$ and $\res{c}$ are strictly increasing in their respective argument.
Thus, there is, at maximum, one unique root within the physically possible domain $\T{h2}^* \in \left[ \T{w2}\, ;\, \T{h1} \right]$ and $\T{c2}^* \in \left[ \T{c1}\, ;\, \T{w1} \right]$.
In general, the root determination of (\ref{eq:resh})--(\ref{eq:resc}) necessitates a numerical multiple-step procedure.
Especially if the included enthalpy calculation is accomplished by real gas models, the overall computation of the reference output model (subscript $r$)
 \begin{align}
\fcn{\underline{g}_r}{\underline{x},\,\underline{u},\,\underline{\theta},\, t} = 
\begin{bmatrix}
\text{root of} \ \res{h}\\ \text{root of} \ \res{c}
\end{bmatrix}
\label{eq:Ausgangsgleichung_Referenzmodell}
\end{align}
comes with a high numerical burden.
For this reason, we will not deploy this reference model $\vec{g}_r$ in the scheme of online monitoring.
Advantageously, this model formulation is the exact description of the reference counterflow process without any further assumptions compared to the list given in Section \ref{sec:ReferenceProcess}.
On that account, the reference model will be used to calculate the time series of the temperatures and thermal properties, on which the validation of the approximate model and the proof of the thermal property `observability' will be accomplished.

\subsubsection{Approximate Model}
\label{sec:OutputEq_approx}

This section concerns simplifying the output equation (\ref{eq:Ausgangsgleichung_Referenzmodell}) so that it becomes solvable in one step without iterations.
Two issues imply the necessity of the multiple-step approach above:
i) the integration of an arbitrarily complex enthalpy calculation model into the root determination problem and
ii) the log-mean temperature difference.

\paragraph{Externalizing the Enthalpy Calculation Model}

\begin{figure*}
	\begin{center}
		\includegraphics{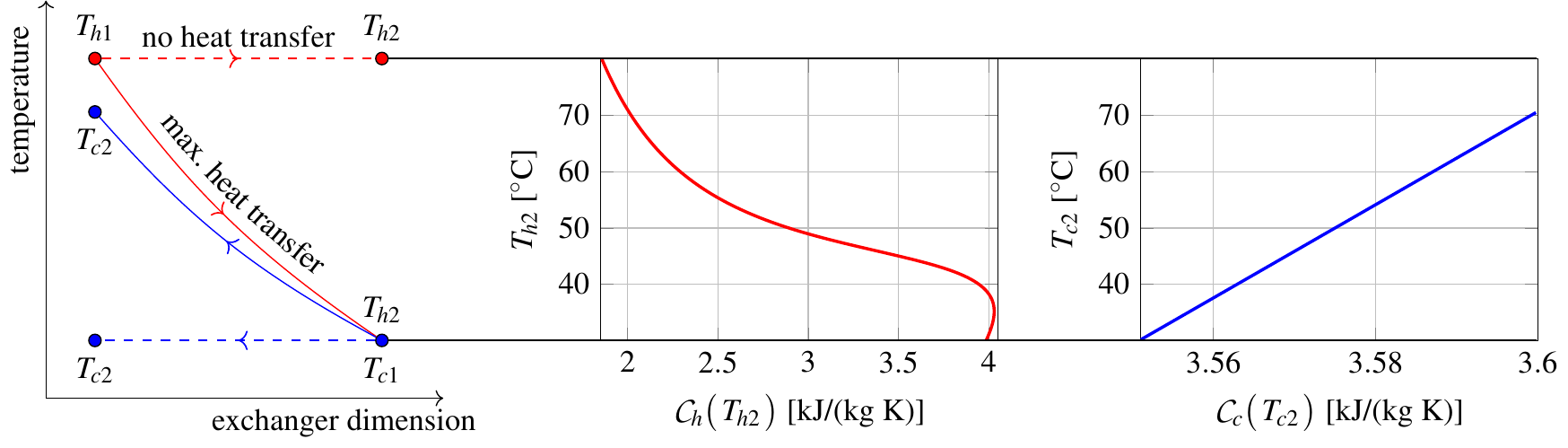}
	\end{center}
	\caption[The mean specific heat range of a supercritical carbon dioxide exchanger]{Mean specific heat range of a supercritical carbon dioxide exchanger; \\ process medium: carbon dioxide with $\md_h = \SI{30}{\kilogram/\second}$, $p_h = \SI{100}{\bar}$; ~ coolant: glycosol-water with $\md_c = \SI{41}{\kilogram/\second}$, $p_c = \SI{4}{\bar}$
	}
	\label{fig:cpMean}
\end{figure*}

To facilitate the one-step-solvable outlet temperature calculation, irrespective of the applied enthalpy calculation model, we introduce mean specific heat parameters
 \begin{align}
\cph &= \fcn{\mathcal{C}_h}{\T{h2}^-} := \frac{ \fcn{h_h}{\T{h2}^-,\, p_h} - \fcn{h_h}{\T{h1},\, p_h} }{ \T{h2}^- - \T{h1}} \, , \label{eq:cph} \\
\cpc &= \fcn{\mathcal{C}_c}{\T{c2}^-} := \frac{ \fcn{h_c}{\T{c2}^-,\, p_c} - \fcn{h_c}{\T{c1},\, p_c} }{ \T{c2}^- - \T{c1}}  \, , \label{eq:cpc}
\end{align}
where $\T{h2}^-$ and $\T{c2}^-$ are the model outputs from a previous time step, and, as above, $h_{h/c}(T, p)$ denotes an enthalpy model, possibly featuring real gas behavior.
Note that $\cph$ and $\cpc$ are time-variant in general, but they are fixed for the root determination step solely, where we replace the enthalpy calculation in (\ref{eq:Hhdot})--(\ref{eq:Hcdot}) with the approximations
\begin{equation}
\fcn{\tilde{h}_{h/c}}{T,\, p} = \theta_{3/4}\cdot T\, .
\label{eq:htilde}
\end{equation}
This is somewhat different from using a calorically perfect gas model in general but manifests the externalization of an unspecified enthalpy calculation model out of the root determination of (\ref{eq:resh})--(\ref{eq:resc}).
To point out the differences to an integrated real-gas model, a specific case study will be considered in Section \ref{sec:steady_transient_monitoring}, for which the dependencies described in Eqs.\ (\ref{eq:cph})--(\ref{eq:cpc}) are depicted in Fig.\ \ref{fig:cpMean}.
The chosen reference will be a supercritical carbon dioxide heat exchanger, 
for which a calorically perfect gas model (constant specific heat) is particularly improper.

\paragraph{Replacing the Logarithmic Mean}

Despite the simplification affecting the calculation of $\mathcal{H}_{h/c}$, the log-mean approach in $\mathcal{Q}$ impedes a one-step-solution.
To find a proper approximation of the logarithmic mean, we look at the arithmetic-logarithmic-geometric mean inequality \cite{nelsen1995proof}:
 \begin{align}
\GM{z_1}{z_2} < \LM{z_1}{z_2} < \AM{z_1}{z_2} \, ,\label{eq:GM_LM_AM_1}
\end{align}
for $\left( z_1,\, z_2 \right) \in \mathbf L_{z_1}^{z_2}$.
Here, 
 \begin{align}
\GM{z_1}{z_2} := \sqrt{z_1\cdot z_2}
\end{align}
denotes the geometric mean.
This relationship motivates the replacement of the logarithmic mean by the following weighted mean $\mathcal{WM}$:
 \begin{align}
\WM{z_1}{z_2} := &\beta\cdot \GM{z_1}{z_2} \nonumber\\
&+ \left(1-\beta\right)\cdot \AM{z_1}{z_2}\, ,
\label{eq:WM}
\end{align}
where $\beta$ is a novel weighting parameter that has to be bounded between $0$ and $1$ if $\mathcal{WM}$ is to serve as a proper substitution for $\mathcal{LM}$ in Eq.\ (\ref{eq:GM_LM_AM_1}).
A proper determination of $\beta$ concerning the domain of the approximate output equations is given in what follows.
\\ \\
For the sake of convenience, we replace the \hl{separated terms for the hot and cold sides with substitutes that cover both sides according to} Table \ref{tab:GLS_Analogie}.
Given that, it is sufficient to \hl{solve the following universal residual and recover the side-specific expressions afterwards:}
 \begin{align}
\fcn{\Rtilde}{\DT{II}^*} &:= \fcn{\Htilde\!}{\DT{II}^*} - \fcn{\Qtilde}{\DT{II}^*} \, ,
\label{eq:GLS2_Start}
\end{align}
where $\DT{II}^* \in \left[ 0\, ;\, \DT{I}+\DT{w} \right]$.
\hl{Here, the recently mentioned approximations, denoted by symbols with $\sim$, are effective and all dependencies can be reformulated in terms of temperature differences $\DT{}$.}
The approximated enthalpy and heat transfer rates are
 \begin{align}%
\fcn{\Htilde\!}{\DT{II}^*} &:= \gamma\cdot C_p\cdot \Big[ \DT{I} - \DT{II}^* + \DT{w} \Big] \, , \label{eq:Htilde}\\
\fcn{\Qtilde}{\DT{II}^*} &:= \gamma\cdot \aA \cdot \WM{\DT{I}}{\DT{II}^*} \, . 
\label{eq:GLS2_Ende}
\end{align}
\begin{table}%
	\centering
	\renewcommand{\arraystretch}{1.5}
	\begin{tabular}{|c|c|c|}
		\hline \hl{substitute} & hot side term & cold side term \\ 
		\hline $\DT{I}$ & $\T{h1}-\T{w1}$ & $\T{w2}-\T{c1}$ \\ 
		\hline $\DT{II}^*$ & $\T{h2}^*-\T{w2}$ & $\T{w1}-\T{c2}^*$ \\ 
		\hline $\DT{w}$ & $\T{w1}-\T{w2}$ & $\T{w1}-\T{w2}$ \\ 
		\hline $C_p$ & $\mcph$  & $\mcpc$ \\ 
		\hline $\gamma$ & -1 & 1 \\ 
		\hline $\aA$ & $\theta_1$ & $\theta_2$ \\ 
		\hline 
	\end{tabular}
	\caption{\hl{Substitutions}}
	\label{tab:GLS_Analogie}
\end{table}%
In contrast to Eqs.\ (\ref{eq:resh})--(\ref{eq:resc}), the root of Eq.\ (\ref{eq:GLS2_Start}) can be calculated directly:
 \begin{align}
\text{root of}\ \Rtilde &= \Fcn{\mathcal{G}}{\DT{I},\,\DT{w},\,\aA,\,C_p,\,\beta} \nonumber\\
&\hspace*{-1.5cm}:=\DT{I} + \DT{w} + \frac{2\, \aA \cdot \beta\cdot \Big[ \DT{I}\cdot\aA\cdot\beta - \xi_4 \Big]}{\xi_1^2}  \nonumber\\
&\hspace*{0.5cm}+ \frac{\aA\cdot \Big[ 2\, \DT{I} + \DT{w} \Big] \cdot \left(\beta - 1\right)}{\xi_1} \, , 	\label{eq:DT2_Loesung}
\end{align}
where we have used the following abbreviations
 \begin{align}
\xi_{1}  &= \aA\cdot\left(1 - \beta\right) + 2\, C_p\, ,\\
\xi_{2}  &= 2\, \aA \cdot \Big[ \aA\cdot\DT{I} - C_p\cdot \DT{w} \Big]\, ,\\
\xi_{3}  &= 4\, C_p^2\cdot \left( \DT{I} + \DT{w} \right) \nonumber\\
&\hspace*{1cm}+ \aA\cdot \Big[ 2\, C_p\cdot \DT{w} - \aA\cdot \DT{I}\Big]\, , \\
\xi_{4}  &= \sqrt{ \left(\xi_2\cdot \beta + \xi_3\right)\cdot \DT{I} }\, .
\end{align}
To guarantee a real-valued solution inside $\left[ 0\, ;\, \DT{I}+\DT{w} \right]$ (cf.\ Eq.\ (\ref{eq:GLS2_Start})) for a given $\DT{I}$, $\beta$ must be chosen within the domain
$\left(\DT{I}, \, \beta \right) \in \mathbf{L_1} \cup \mathbf{L_2}\,$, where
 \begin{align}
\mathbf{L_1}&= \mathbb{R} \times \{ 0 \} \, , \qquad 
\mathbf{L_2}= \Big(\mathbb R_+ \setminus \{ 0 \} \Big) \times \mathbf{B}  \, ,\\
\mathbf{B} &= \left] 0\, ; \ 1\right] \ \cap\ \Big\{ \beta \,\big|\ \xi_2\cdot \beta + \xi_3 \ge 0\Big\} \, \nonumber\\
&\hspace*{1cm} \cap\ \Big\{ \beta \,\big|\ \DT{I}\cdot\aA\cdot \beta - \xi_4 \le 0 \Big\}\, .
\label{eq:beta_Loesungsraum}
\end{align}
Inside the partial domain $\mathbf{L_1}$, where $\beta=0$, the mean temperature difference is calculated with the arithmetic-mean approach (cf.\ Eq.\ (\ref{eq:WM})), which is the equivalent domain extension, as can be found in Eq.\ (\ref{eq:Qd_unrestricted}).
For most operating conditions, we found that $\mathbf{B}=\left] 0\, ; \ 1\right]$, and, thus, the choice of $\beta$ is not very restrictive.
However, to meet the above specified requirements for all possible operating conditions, we have to think of situations where $\mathbf{B} \subsetneq \left] 0\, ; \ 1\right]$, and the arithmetic-mean approach is not favorable. 
A generally applicable suggestion on how to choose $\beta$ is:

\vspace{0.5em}
\fbox{\parbox{\dimexpr \linewidth - 2\fboxrule - 2\fboxsep - 2\parindent}{If \mbox{$\DT{I}>0$} and \mbox{$\mathbf{B}\ne \emptyset$}, then choose the element of \mbox{$\Big\{ \beta_{LM},\, \beta_1^*,\, \beta_2^* \Big\} \cap \mathbf{B}$} that is the nearest neighbor of $\beta_{LM}$; otherwise, choose \mbox{$\beta=0$}.\footnotemark
}}
\vspace{0.5em}

\footnotetext{
	If $\mathbf{B} \subsetneq \left] 0\, ; \ 1\right]$ is not empty, then \mbox{$\mathbf{B} = \left] 0\, ; \ \beta_{1/2}^*\right]$}, \mbox{$\mathbf{B} = \left[ \beta_{1/2}^*\, ;\ 1 \right]$}, or \mbox{$\mathbf{B} = \left[ \beta_{1/2}^*\, ; \ \beta_{2/1}^* \right]$}.
}%

\noindent
The included terms are calculated as follows:
 \begin{align}
\beta_{LM} &= \frac{\AM{\DT{Is}}{\DT{IIs}} - \LM{\DT{Is}}{\DT{IIs}}}{\AM{\DT{Is}}{\DT{IIs}} - \GM{\DT{Is}}{\DT{IIs}}} \, , \label{eq:betaLM}\\
\beta^*_{1/2} &= \frac{\xi_2 \pm \sqrt{ 4\, \DT{I}\cdot \xi_3 \cdot \aA^2 + \xi_2^2 }}{2\, \DT{I}\cdot \aA^2} \, .
\end{align}
Subscript $s$ denotes the steady-state, the calculation of which is presented in Section \ref{sec:steady}.
For steady-state conditions, the favorable $\beta_{LM}$ yields an exact approximation of the log-mean temperature difference.
Note that one has to determine $\DT{II}^*$ and, thus, $\beta$ for the hot and cold sides, respectively, by substituting the general terms according to Table \ref{tab:GLS_Analogie}.
Finally, in terms of Eq.\ (\ref{eq:Ausgangsgleichung}), the derived one-step-solvable output equations of the approximate model are:
 \begin{align}
&\fcn{\vec{g}}{\underline{x},\,\underline{u},\,\underline{\theta},\, t} \label{eq:Ausgangsgleichung_Ersatzmodell}\\
&\hspace*{0.1cm}= 
\begin{bmatrix}
\Fcn{\mathcal{G}}{\T{h1}-\T{w1},\, \T{w1}-\T{w2},\,\theta_1,\,\md_h\cdot \theta_3,\,\beta_h} + \T{w2} \\ 
\T{w1} - \Fcn{\mathcal{G}}{\T{w2}-\T{c1},\,\T{w1}-\T{w2},\,\theta_2,\,\md_c\cdot \theta_4,\,\beta_c}
\end{bmatrix} \, .
\nonumber
\end{align}

\subsection{Steady State}
\label{sec:steady}

The steady-state is of particular importance in the presented modeling scheme.
It is essential for the determination of Eq.\ (\ref{eq:betaLM}) and the derivation of the dynamic state equations, as presented in Section \ref{sec:ZDGL}.
Within this section, the \textit{a priori} calculation of the steady-state is presented.
For a discussion on the presupposed uniqueness of that steady-state, the reader is referred to the appendices.
The reference model is introduced first in Section \ref{sec:steady_reference} before an approximate solution is derived in Section \ref{sec:steady_approx}.

\subsubsection{Reference Model}
\label{sec:steady_reference}
To obtain the steady-state outlet temperatures $\left(\T{h2s},\, \T{c2s}\right)$ without specifying a simple enthalpy calculation model, the root of the following residual functions has to be determined
\begin{numcases}{}
\fcn{\res{s1}}{\T{h2s}^*,\,\T{c2s}^*} := \fcn{\mathcal{H}_{c}}{\T{c2s}^*} + \fcn{\mathcal{H}_{h}}{\T{h2s}^*} \, , & \label{eq:res_s1} \\
\fcn{\res{s2}}{\T{h2s}^*,\,\T{c2s}^*} := \fcn{\mathcal{H}_{c}}{\T{c2s}^*} & \label{eq:res_s2} \\
\hspace*{1.8cm} - \fcn{\mathcal{Q}}{\T{h1}-\T{c2s}^*,\,\T{h2s}^*-\T{c1},\,\kA} \, ,&\nonumber 
\end{numcases}
applying a numerical multiple-step procedure.
The root determination of (\ref{eq:res_s1})--(\ref{eq:res_s2}) is a mathematical formulation stating the equality of steady heat transfer and enthalpy rates.
In contrast to (\ref{eq:Ausgangsgleichung_Referenzmodell}), this is a coupled problem.
Nevertheless, the root of (\ref{eq:res_s1})--(\ref{eq:res_s2}), which is $\left(\T{h2s},\, \T{c2s}\right)$, is unique.
This is expanded upon in Appendix A.

Further, we have to determine the steady-state values of the wall temperatures $\left(\T{w1s},\, \T{w2s}\right)$, obtained as the root of
\begin{numcases}{}
\fcn{\res{s3}}{\T{w1s}^*,\,\T{w2s}^*} &\label{eq:res_s3}\\
\hspace*{1.0cm}:= \fcn{\mathcal{Q}}{\T{h1}-\T{c2s},\,\T{h2s}-\T{c1},\,\kA} &\nonumber \\
\hspace*{1.2cm} - \fcn{\mathcal{Q}}{\T{h1}-\T{w1s}^*,\,\T{h2s}-\T{w2s}^*,\,\aAh} \, ,&\nonumber \\
\fcn{\res{s4}}{\T{w1s}^*,\,\T{w2s}^*} &\label{eq:res_s4} \\
\hspace*{1.0cm}:= \fcn{\mathcal{Q}}{\T{h1}-\T{c2s},\,\T{h2s}-\T{c1},\,\kA} &\nonumber \\
\hspace*{1.2cm} - \fcn{\mathcal{Q}}{\T{w1s}^*-\T{c2s},\,\T{w2s}^*-\T{c1},\,\aAc} \, .&\nonumber
\end{numcases}
Equations (\ref{eq:res_s3})--(\ref{eq:res_s4}) state the balance between wall heating and cooling fluxes and the total heat transfer, calculated with the above determined steady outlet temperatures.
Again, a unique solution is obtained.
For a sketch of the proof, see Appendix B.
The solution is given by
 \begin{align}
\T{w1s} &= \T{h1} + \frac{\aAc}{\aAh+\aAc}\cdot \left(\T{c2s}-\T{h1}\right)\, ,  \label{eq:Tw1s_aAh}\\
\T{w2s} &= \T{h2s}+ \frac{\aAc}{\aAh+\aAc}\cdot \left(\T{c1}-\T{h2s}\right)\, .  \label{eq:Tw2s_aAh}
\end{align}
Thus, if $\left(\T{h2s},\, \T{c2s}\right)$ are determined, the steady wall temperatures can be calculated directly without a further multiple-step root determination.

\subsubsection{Approximate Model}
\label{sec:steady_approx}
Applying the same strategy here as presented in Section \ref{sec:OutputEq_approx}, i.e., fixing fluid properties concerning a previous time step (superscript $^-$), which led to the simple enthalpy calculation (\ref{eq:htilde}), the steady-state calculation becomes solvable in one step:
 \begin{align}
\T{h2s} &= 
\begin{cases}
\T{c1} + \frac{\displaystyle\left[\T{c1} - \T{h1}\right] \left[\mcpcsND - \mcphsND\right]}{\displaystyle \mcphsND - \mcpcsND \xi_s} & ,\ \frac{\displaystyle \mcphsND}{\displaystyle \mcpcsND}\ne 1 \\[.5em]
\frac{\displaystyle \T{c1}\kA + \T{h1} \mcphsND}{\displaystyle \kA + \mcphsND} & , \ \frac{\displaystyle \mcphsND}{\displaystyle \mcpcsND} = 1
\end{cases} \, ,\label{eq:Th2s_Ersatz_a} \\[.5em]
\T{c2s} &= \T{c1} + \frac{\mcphsND}{\mcpcsND}\left[ \T{h1} - \T{h2s} \right] \, , \label{eq:Tc2s_Ersatz_a}
\end{align}
where
 \begin{align}
\xi_s &= \exp \left( \frac{\kA}{\mcphsND} - \frac{\kA}{\mcpcsND} \right) \, ,\\
\theta_5 &= \fcn{\mathcal{C}_h}{\T{h2s}^-} \, , \qquad 	\theta_6 = \fcn{\mathcal{C}_c}{\T{c2s}^-}  \, .\label{eq:cpcs}
\end{align}
Clearly, the steady outlet temperatures of the approximate model given the fixed inputs and parameters are unique according to (\ref{eq:Th2s_Ersatz_a})--(\ref{eq:Tc2s_Ersatz_a}).
Likewise, this holds true for the steady-state calculated with Eqs.\ (\ref{eq:Tw1s_aAh})--(\ref{eq:Tw2s_aAh}).
The aforementioned proof of uniqueness is valid for the approximate model as well because the enthalpy calculation model was not specified further, and, thus, the simplified enthalpy rate equation of the approximate model (cf.\ Eq.\ (\ref{eq:Htilde})) is already included.

Note that the log-mean approximation does not affect the steady-state calculation if $\beta=\beta_{LM}$ is set according to Eq. (\ref{eq:betaLM}).
If so, the steady log-mean temperature difference equals the steady weighted-mean temperature difference.

\subsection{Dynamic State Equations}
\label{sec:ZDGL}

To completely specify the model (\ref{eq:ZDGL})--(\ref{eq:Ausgangsgleichung}), the right hand side of Eq.\ (\ref{eq:ZDGL}) has to be known, i.e., a dynamic model is needed to describe the temporal evolution of the two wall temperatures, $\T{w1}(t)$ and $\T{w2}(t)$.
The steady-state values $\T{w1s}$ and $\T{w2s}$ were already determined in Section \ref{sec:steady}.
$\T{w1}(t)$ and $\T{w2}(t)$ change as functions of time $t$ due to driving temperature differences concerning the process and cooling media.
However, in an attempt to keep the model order as low as possible, dynamic states for the temperatures of the hot and cold fluids were discarded.
Therefore, a black-box-like approach was chosen.

\begin{figure}
	\begin{center}
		\includegraphics{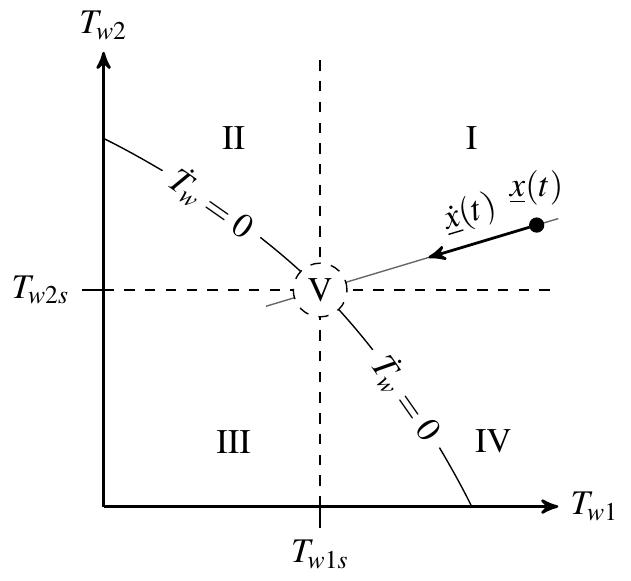}
	\end{center}
	\caption{The state space}
	\label{fig:StateSpace}
\end{figure}

As the system is stable, $\T{w1}(t)$ and $\T{w2}(t)$ tend towards $\T{w1s}$ and $\T{w2s}$ (see Fig.\  \ref{fig:StateSpace}), which are unique values given fixed input and parameter vectors (see Section \ref{sec:steady} and appendices).
As an oscillatory-like approach of the steady-state is highly unlikely, it is proposed that the gradient points directly to the steady state.
This can be described by
\vspace{1em}
 \begin{align}
\begin{bmatrix}
\smallfcn{\Td{w1}}{t} \\ \smallfcn{\Td{w2}}{t}
\end{bmatrix} = a\cdot 
\begin{bmatrix}
\smash{\overbrace{\smallfcn{\T{w1s}}{\vec{\theta},\vec{u},t} - \smallfcn{\T{w1}}{t}}^{\smallfcn{\ew{1}}{\vec{x},\vec{\theta},\vec{u},t}}} \\ \smash{\underbrace{\smallfcn{\T{w2s}}{\vec{\theta},\vec{u},t} - \smallfcn{\T{w2}}{t}}_{\smallfcn{\ew{2}}{\vec{x},\vec{\theta},\vec{u},t}}}
\end{bmatrix}
\quad ,\ a \in \mathbb{R_+}\, .
\label{eq:Tw_a_ew}
\end{align}
Despite its black-box character, a physical constraint has to be met for Eq.\ (\ref{eq:Tw_a_ew}).
The temperature rates, which are determined by a common parameter $a$, will depend on the total heat capacity of the wall $\theta_7$ and on the heat fluxes, $\Qd_h$ and $\Qd_c$.
An overall balance of the wall results in
 \begin{align}
\theta_7\cdot \Td{w} = - \Qd_h - \Qd_c \label{eq:lumpedCapacity}\, .
\end{align}
$\Qd_h$ and $\Qd_c$ can be determined according to Eqs.\ (\ref{eq:Qd_extended})--(\ref{eq:Qc_extended}) with the respective outlet temperatures of either the reference or the approximate model.
Further, as the parameter $\theta_7$ is either known from first principles or from an identification using historical data, Eq.\ (\ref{eq:lumpedCapacity}) can be utilized to determine a mean temperature change $\Td{w}$ that has to be distributed to $\Td{w1}$ and $\Td{w2}$ by the proper choice of the parameter $a$.
To this end, we first look at a combination of Eq.\ (\ref{eq:Tw_a_ew}) and Eq.\ (\ref{eq:lumpedCapacity}).

An intuitive way would be to assume that the average temperature change of the refined model equals the temperature change of the lumped one, i.e.,
 \begin{align}
\Td{w} = \AM{\Td{w1}}{\Td{w2}} \, . \label{eq:Tdw_I_III}
\end{align}
This yields reasonable norm values of $\vec{\dot{x}}$ within sectors I and III, where both wall temperatures either increase or decrease (cf.\ Fig.\ \ref{fig:StateSpace}), bounded according to $\lVert \vec{\dot{x}} \rVert < 2\, \left| \Td{w} \right|$ inside the respective area and with $\lVert \vec{\dot{x}} \rVert = 2\, \left| \Td{w} \right|$ on the borderlines of sectors II and IV.
On the contrary, due to the opposite signs of $\Td{w1}$ and $\Td{w2}$, the norm values are unlimited within II and IV if Eq.\ (\ref{eq:Tdw_I_III}) is forced.
To overcome this issue with a physically reasonable behavior, we postulate $\lVert \vec{\dot{x}} \rVert = 2\, \left| \Td{w} \right|$ within sectors II and IV.
Finally, the overall norm design in this black-box model is achieved by setting
 \begin{align}
a = 
\begin{cases}
\frac{2\,\Td{w}}{\ew{1}+\ew{2}} & \text{within sectors I and III} \, ,\\[.5em]
\frac{2\, \left| \Td{w} \right|}{\sqrt{\ew{1}^2 + \ew{2}^2}} & \text{within sectors II and IV, and}\\[.5em]
0 & \text{within sector V}\, .
\end{cases}
\end{align}
Note that sector V has been introduced for numerical reasons. It is finite but arbitrarily small concerning numerical accuracy.
Further, we suggest setting a lower bound for $\left|\Td{w}\right|$ within sectors II and IV to guarantee asymptotic stability.
Otherwise, the model could, metaphorically speaking, become stuck at the \mbox{``$\Td{w}=0$ graph"} in Fig.\ \ref{fig:StateSpace}, which is a curved line in the state space, depicting the set of steady-states of the lumped capacity model.\footnote{
	Referring to the remarks in Section \ref{sec:steady}, this line must contain the unique steady-state $\left(\T{w1s},\, \T{w2s}\right)$.}

\section{Monitoring}
\label{sec:monitoring}

The proposed model-based monitoring scheme will consist of a real-time estimation of parameters of the approximate model introduced above.
A proper parametrization will be presented in Section \ref{sec:parameters}.
The real-time estimation is done in the framework of the Extended Kalman-Filter, summarized in Section \ref{sec:estimation}.
Then, Section \ref{sec:steady_transient_monitoring} shows the application of the monitoring scheme in a steady-state as well as in highly transient operating conditions.
In Section \ref{sec:reduced_information}, we address a specific situation in which information, provided for the monitoring algorithm, is uncertain or missing.

\subsection{Parametrization}
\label{sec:parameters}

The literature is full of empirically derived correlations between the overall convective heat transfer coefficient and the operating conditions expressed with dimensionless numbers (e.g., \cite{baehr1996warme}).
We want to offer the option to partly integrate such dependencies into the monitoring scheme, as we address flexible plants that will often run in non-steady-state operations during which the heat transfer coefficients might change dynamically.
All of the approaches can be reformulated as
 \begin{align}
\text{Nu}_m = c_1\cdot \text{Re}_m^{E_1} \cdot \text{Pr}_m^{E_2}\cdot  \fcn{f_\alpha}{\text{Re}_m,\,\text{Pr}_m} + c_2 \, ,
\label{eq:Num}
\end{align}
where $c_i$, $E_j$, and $f_\alpha$ denote coefficients, exponents, and a function depending on the specific approach, respectively, and $\text{Nu}_m$, $\text{Re}_m$, and $\text{Pr}_m$ are the well-known dimensionless numbers; more precisely, they are
 \begin{flalign}
&\text{the overall Nusselt number} &\text{Nu}_m &= \frac{\alpha_m\cdot L}{\lambda_m}\,,&&&\\
&\text{the overall Reynolds number} & \text{Re}_m &= \frac{w_m\cdot \rho_m \cdot L}{\eta_m}\,,\\
&\text{the overall Prandtl number} & \text{Pr}_m &= \frac{\eta_m\cdot c_{pm}}{\lambda_m}\, .
\end{flalign}
Here, $w_m$ and $L$ denote the mean fluid velocity and a reference length, respectively. The included fluid properties, such as density $\rho_m$, viscosity $\eta_m$, specific heat $c_{pm}$, and thermal conductivity $\lambda_m$, are typically evaluated based on the arithmetic mean of the intake and outlet temperatures. 

In contrast to the reference model, which accounts for arbitrarily complex approaches in the manner of Eq. (\ref{eq:Num}), we do not intend to presuppose the existence of such a general fluid property model within the monitoring scheme that is based on the approximative model.
A simpler correlation is supposed instead.
Discarding $f_\alpha$ and the temperature dependencies of the fluid properties
motivates the following approach:\footnote{
	Physical dimensions: $\left[\upsilon_{h/c}\right]=\left[\theta_{h3/c3}\right]=\SI{}{\watt/\kelvin}$,  $\left[\md\right]=\SI{}{\kilogram/\second}$, and $\left[\bar{c}_p\right]=\SI{}{\joule/(\kilogram\, \kelvin)}$
}
 \begin{align}
\aAh &= \upsilon_h  \left(\frac{\md_h}{\SI{1}{\kilogram/\second}}\right)^{\theta_{h1}} \left(\frac{\bar{c}_{ph}}{\SI{1}{\joule/(\kilogram\, \kelvin)}}\right)^{\theta_{h2}} + \theta_{h3} \, ,
\label{eq:aAh_Ansatz}\\
\aAc &= \upsilon_c \left(\frac{\md_c}{\SI{1}{\kilogram/\second}}\right)^{\theta_{c1}} \left(\frac{\bar{c}_{pc}}{\SI{1}{\joule/(\kilogram\, \kelvin)}}\right)^{\theta_{c2}} + \theta_{c3} \, ,
\label{eq:aAc_Ansatz}
\end{align}
where we set $\left(\bar{c}_{ph},\, \bar{c}_{pc}\right)=\left(\theta_5,\,\theta_6\right)$ in the scope of the \textit{a priori} steady-state calculation and $\left(\bar{c}_{ph},\, \bar{c}_{pc}\right)=\left(\theta_3,\,\theta_4\right)$ otherwise.
Here, $\p = \begin{bmatrix}\upsilon_h & \upsilon_c\end{bmatrix}^T$ are time-variant parameters that will be estimated in the context of monitoring, and $\vec{\theta}_{hc} = \begin{bmatrix}\theta_{h1\dots 3} & \theta_{c1\dots 3}\end{bmatrix}^T$ are time-invariant model parameters, which should be identified using the historical data of the individual heat exchanger.
If there are no proper data, $\vec{\theta}_{hc}=\vec{0}$ yields the primary approach $\p = \begin{bmatrix}\aAh& \aAc\end{bmatrix}^T$.
Remember that slow varying model parameters are assumed in general.
The better this assumption holds, the better the model will perform.
That means, e.g., if a relevant correlation between $\aAh$ and $\md_h$ is known, it would be unreasonable to set $\theta_{h1}=0$.
The model-based estimator introduced below would then perform worse in tracking $\aAh$ during phases of fast varying $\md_h$.

\subsection{Joint Estimation}
\label{sec:estimation}

For the joint estimation of model states and the $\p$-parameters, the well-known Extended Kalman Filter (EKF) scheme is applied. It is referred to as the Joint-EKF approach by the state estimation community.
Here, for online monitoring, a real-time estimation of $\kA$ is accomplished.
The EKF is a recursively formulated model-based estimation method, and, thus, it is an eligible online estimator.
It assumes a stochastic system formulation in the sense of
\begin{numcases}{\hspace*{-0.75cm}}
\smallfcn{\vec{\dot{x}}_\upsilon}{t} \sim \Fcn{\mathcal{N}}{ \fcn{\vec{f}_\upsilon}{\vec{x}_\upsilon,\vec{u},\vec{\theta},t},\, \mat{R_{x\upsilon}}}\, , &\hspace*{-0.6cm} $\smallfcn{\vec{x}_\upsilon}{t_0} = \vec{x}_{\upsilon 0}\, ,$ \label{eq:Nxv}\\
\smallfcn{\vec{y}}{t} \sim \Fcn{\mathcal{N}}{ \fcn{\vec{g}_\upsilon}{\vec{x}_\upsilon, \vec{u}, \vec{\theta}, t},\, \mat{R_y}} \, , \label{eq:Ny}
\end{numcases}
where \hl{the expression} $\smallfcn{\vec{z}}{t}\sim \fcn{\mathcal{N}}{\smallfcn{\vec{z}_m}{t},\mat{R_z}}$ \hl{denotes representatively that a} vector $\vec{z}$ is normally distributed with the mean $\vec{z}_m$, and $\smallfcn{\vec{z}}{t}-\smallfcn{\vec{z}_m}{t}$ is a continuous white noise process with a spectral density matrix $\mat{R_z}$.
To satisfy the postulated demand, the joint estimation requires
 \begin{align}
\vec{x}_\upsilon &= \begin{bmatrix}\vec{x}\\ \vec{\upsilon}\end{bmatrix} \, , & \fcn{\vec{f}_\upsilon}{\vec{x}_\upsilon, \vec{u}, \vec{\theta}, t} &= \begin{bmatrix}\fcn{\vec{f}}{\vec{x}, \vec{u}, \vec{\theta}_\upsilon, t}\\\vec{0}\end{bmatrix} \, , \label{eq:joint_model}\\
\mathbf{R_{x \upsilon}} &= \begin{bmatrix}\mathbf{R_x} & \mathbf{0} \\\mathbf{0} &   \mathbf{R_\upsilon}\end{bmatrix} \, , &\fcn{\vec{g}_\upsilon}{\vec{x}_\upsilon, \vec{u}, \vec{\theta}, t} &= \fcn{\vec{g}}{\vec{x}, \vec{u}, \vec{\theta}_\upsilon, t}\, , \\[.5em]
\vec{\theta}_\upsilon &= \begin{bmatrix}\smallfcn{\theta_1}{\p,\vec{\theta}_{hc}}~ & \smallfcn{\theta_2}{\p,\vec{\theta}_{hc}}~ & \theta_3~ & \dotsm & ~\theta_7\end{bmatrix}^T\, ,  \hspace*{-6.5cm}
\end{align}
where $\smallfcn{\theta_1}{\p,\vec{\theta}_{hc}}$ and $\smallfcn{\theta_2}{\p,\vec{\theta}_{hc}}$ are the overall convection conductances calculated with Eqs.\ (\ref{eq:aAh_Ansatz}) and (\ref{eq:aAc_Ansatz}), respectively.
As seen in Eq.\ (\ref{eq:joint_model}), the approximate model $\left(\vec{f},\, \vec{g}\right)$ is part of the model used in the Joint-EKF.
Although the spectral density matrices $\mat{R_x}$, $\mat{R_\upsilon}$, and $\mat{R_y}$ are well-defined by Eqs.\ (\ref{eq:Nxv})--(\ref{eq:Ny}), they are typically unknown by value for a specific application.
Therefore, they are interpreted as tunable design parameters of the Joint-EKF approach.
In a common application and for the presented scenarios below, measurements $\vec{y}$ are not accessible in a continuous manner but at discrete points in time \hl{$t_k$.} 
On that account, the Joint-EKF is implemented with a time-continuous prediction step
\begin{align}
&\mathcolorbox{yellow}{
	\vec{\hat{x}}^-_\upsilon(t)=\vec{\hat{x}}_\upsilon(t_{k-1}) + \int\limits_{t_{k-1}}^{t} \fcn{\vec{f}_\upsilon}{\vec{\hat{x}}^-_\upsilon, \vec{u}, \vec{\theta}, \tau} \text{d}\tau\, ,}\\
\begin{split}
&\mathcolorbox{yellow}{\mathbf{P}^-(t) = \mathbf{P}(t_{k-1}) + \mathbf{R_{x \upsilon}}\cdot \left(t-t_{k-1}\right) }\\
&\mathcolorbox{yellow}{\hspace*{1cm}+ \int\limits_{t_{k-1}}^{t} \left( \mathbf{F}(\tau)\mathbf{P}^-(\tau) + \mathbf{P}^-(\tau)\mathbf{F}(\tau)^T  \right) \text{d}\tau}
\end{split}
\end{align}
\hl{within the time interval $t \in \left[t_{k-1};\, t_k\right]$} and a time-discrete measurement update
\begin{align}
&\mathcolorbox{yellow}{\mathbf{K} = \mathbf{P}^-(t_k) \mathbf{H}^T \left( \mathbf{H}\mathbf{P}^-(t_k)\mathbf{H}^T + \mathbf{R}_y\cdot \left(t_k-t_{k-1}\right)^{-1} \right)^{-1}\, ,}\\
&\mathcolorbox{yellow}{\vec{\hat{x}}_\upsilon(t_k) = \vec{\hat{x}}_\upsilon^-(t_k) + \mathbf{K} \left(\vec{y}(t_k) - \fcn{\vec{g}_\upsilon}{\vec{\hat{x}}^-_\upsilon, \vec{u}, \vec{\theta}, t_k}\right)\, ,}\\
&\mathcolorbox{yellow}{\mathbf{P}(t_k) = \mathbf{P}^-(t_k) - \mathbf{K}\mathbf{H}\mathbf{P}^-(t_k)\, .}
\end{align}
\hl{The introduced jacobian matrices}
\begin{align}
&\mathcolorbox{yellow}{\mathbf{F}(t) := \frac{\partial}{\partial\vec{x}_\upsilon}\fcn{\vec{f}_\upsilon}{\vec{x}_\upsilon, \vec{u}, \vec{\theta}, t}\Big\vert_{\vec{x}_\upsilon=\vec{\hat{x}}_\upsilon^-(t)}}\\
&\mathcolorbox{yellow}{\mathbf{H} := \frac{\partial}{\partial\vec{x}_\upsilon}\fcn{\vec{g}_\upsilon}{\vec{x}_\upsilon, \vec{u}, \vec{\theta}, t_k}\Big\vert_{\vec{x}_\upsilon=\vec{\hat{x}}_\upsilon^-(t_k)}}
\end{align}
\hl{can be analytically derived for the approximate model, which is a further advantage of this model.
The ``\textasciicircum'' symbol is used to distinguish the estimate }\colorbox{yellow}{$\vec{\hat{x}}_\upsilon$}\hl{ from the true state }\colorbox{yellow}{$\vec{x}_\upsilon$}\hl{, which is unknown in a real experiment.
Matrix $\mathbf{P}$ serves as an estimate for the covariance matrix of the error between estimated and true state.
For further details on the EKF, the reader is referred to \mbox{\cite{gelb1974applied}}.
For the experiments below, we set biased start conditions }\colorbox{yellow}{$\vec{\hat{x}}_\upsilon(t_0=\SI{0}{\minute})$ and, further, $\mathbf{P}(t_0=\SI{0}{\minute})=\SI{1}{\second}\cdot \mathbf{R_{x\upsilon}}$}\hl{ to initialize the algorithm.
}

\subsection{Monitoring in Steady and Dynamic Operating Conditions}
\label{sec:steady_transient_monitoring}

\begin{figure}
	\begin{center}
		\fcolorbox{yellow}{yellow}{\includegraphics[scale=1]{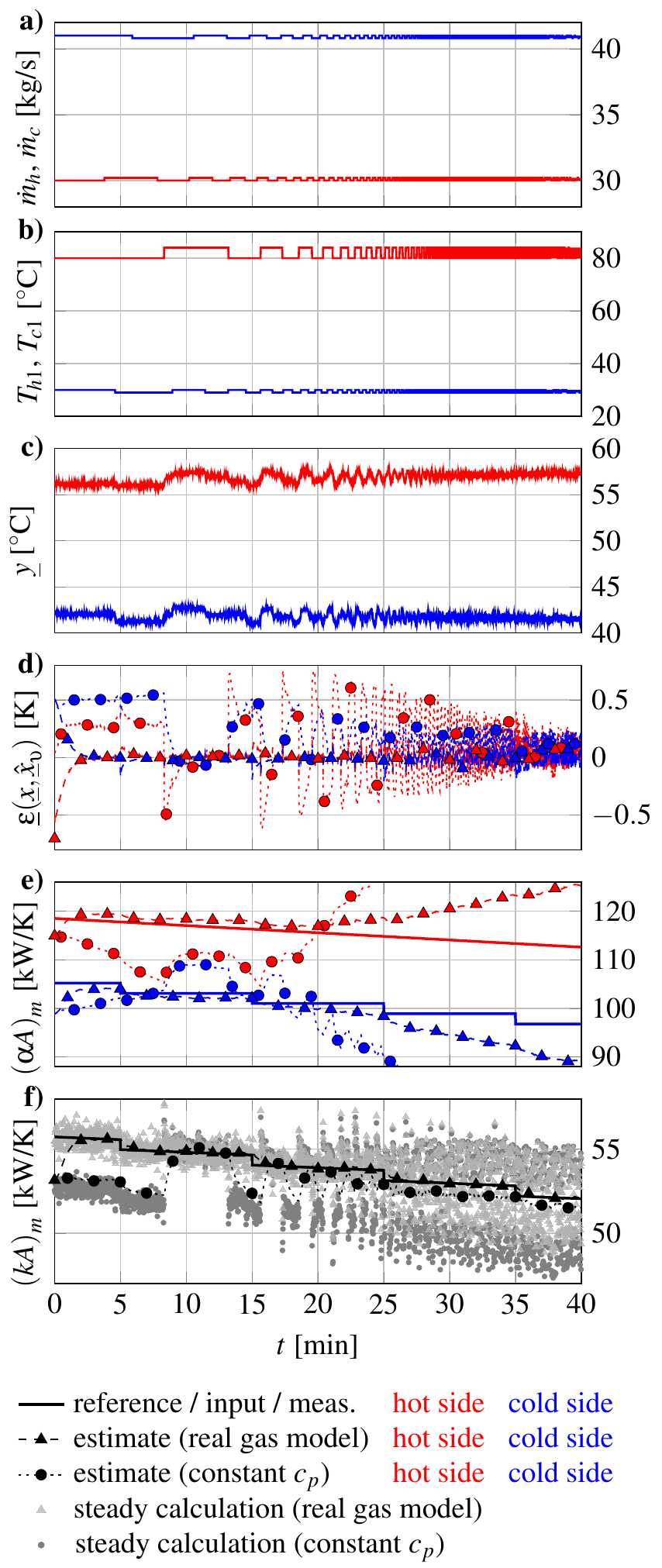}}
	\end{center}
	\caption[Monitoring from steady to high-frequency operating conditions]{Model-based and model-free monitoring from steady (\mbox{$t=\SI{0}{min}$}) to high-frequency (\mbox{$t\rightarrow \SI{40}{min}$}) operating conditions;\\ a) preset mass flows; b) preset intake temperatures; c) noisy measurements; d) noise canceled deviation \mbox{$\smallfcn{\vec{\epsilon}}{\vec{x}, \vec{\hat{x}}_\upsilon} = \smallfcn{\vec{g}_r}{\vec{x},\vec{u},\vec{\theta},t} - \smallfcn{\vec{g}_\upsilon}{\vec{\hat{x}}_\upsilon,\vec{u},\vec{\theta},t}$}; e) preset and estimated overall convection conductances; and f) preset, estimated, and model-free calculated overall thermal conductance
	}
	\label{fig:varObsv}
\end{figure}

In this section, on the basis of a simulation experiment, the advantage of the suggested monitoring scheme over a conventional (model-free) thermal rating is noted, where the latter is more precisely the \textit{steady calculation} of $\kA$ according to Eq.\ (\ref{eq:kA_stat}).
Furthermore, the influence of the applied enthalpy calculation model is discussed herein.
The relevant time series, which belong to the considered experiment, are depicted in Fig.\ \ref{fig:varObsv}.

The experimental setup is as follows:
Within the reference model, a real gas model \cite{lemmon2010nist,span1996new} is used to calculate specific enthalpies of the carbon dioxide process fluid at a supercritical state.
The coolant is a glycosol-water-mixture that is described by a thermally perfect fluid model ($d\smallfcn{h_c}{T}=\smallfcn{c_{pc}}{T}dT$; data from \cite{glykosol}).
The preset temporal variations of $\smallfcn{\aAh}{t}$ and $\smallfcn{\aAc}{t}$ are denoted as \textit{reference} trends within Fig.\ \ref{fig:varObsv}\,e.
Their combination, according to Eq.\ (\ref{eq:kA_dyn}), yields the \textit{reference} trend $\smallfcn{\kA}{t}$\hl{, which is depicted in \mbox{Fig.\ \ref{fig:varObsv}\,f.}}
The thermal rating is stated as successful if an algorithm is capable of tracking those reference trends on the basis of the system inputs $\vec{u}$ and measured outlet temperatures $\vec{y}$.
Here, the measurements (Fig.\ \ref{fig:varObsv}\,c) are the superposition of the reference model's output $\smallfcn{\vec{g}_r}{\vec{x},\vec{u},\vec{\theta},t}$ and an artificial, normally distributed noise with a standard deviation of \SI{0.1}{K}.
The inputs \hl{\mbox{(Fig.\ \ref{fig:varObsv}\,a--b)}} are chosen such that the simulated exchanger continuously runs from a steady operating point at $t=\SI{0}{min}$ to the measurable highest-frequent transient phase at $t=\SI{40}{min}$ (test frequency: $\Delta t^{-1}=\SI{1}{Hz}$).
Quite clearly, this is an unrealistic but eligible excitation to demonstrate the functionality of the monitoring scheme.
The settings for the monitoring algorithm are:
 \begin{align} \begin{split}
&\mathbf{R_x} = \SI{0.1}{\second}\cdot \left( \frac{\Qd_{design}}{100\cdot \theta_7} \right)^2 \cdot \mathbf{I}_2\, , \quad \mathcolorbox{yellow}{\mathbf{R_y} = \SI{1}{\second}\cdot\left(\SI{0.1}{\kelvin}\right)^2\cdot \mathbf{I}_2\, ,} \\
&\mathbf{R_\upsilon} = \SI{0.1}{\second}\cdot \left(\SI{100}{\watt/(\kelvin\,\second)}\right)^2 \cdot \mathbf{I}_2\, , \quad \vec{\theta}_{hc} = \vec{0}\, , \label{eq:Rp}
\end{split}\end{align}
where $\Qd_{design} = \SI{1.6}{\mega\watt}$ and $\theta_7 = \SI{566.5}{\kilo\joule/\kelvin}$ are first principle parameters of the specific exchanger, and $\mathbf{I}_2$ represents the $2\times 2$ identity matrix.

Without model-based estimation, one could calculate the overall thermal conductance from the inputs and measurements directly if the exchanger's dynamic is ignored.
Typically, the hot-side enthalpy rate would be preferred (cf.\ Eq.\ (\ref{eq:Hhdot})) to replace the total heat transfer rate in Eq.\ (\ref{eq:kA_stat}) because of the smaller impact of potential measurement errors.
Such rating points are depicted within Fig.\ \ref{fig:varObsv}\,f for two cases: i) if the (correct) real gas enthalpy calculation is applied and ii) if calorically perfect gas behavior is assumed (\textit{constant} $c_p$).
Despite an adequate choice for the constant caloric heat value\footnote{
	With respect to the caloric dependance, according to Fig.\ \ref{fig:cpMean} and the range of measured outlet temperatures, we set \mbox{$\smallfcn{h_h}{T, p}= \SI{2.3}{kJ/(kg\ K)} \cdot T$} for the \textit{constant} $c_p$ approach.
},
this fluid model yields biased rating points, where the bias depends on the operating point, which is undesirable for the flexible monitoring task.

Even if the (bias free) real gas model is applied in the approach without the Joint-EKF, two effects are superposed: i) the (temporally constant) measurement noise impact and ii) the (temporally growing) impact of the ignored dynamic, causing the cumulative outliers (see Fig.\ \ref{fig:varObsv}\,f).
In this respect, the Joint-EKF serves as a noise filter that is insensitive to the exchanger's dynamic behavior.
That is why the filtering performance is quite stable for the whole experiment\hl{\mbox{, cf.\ Fig.\ \ref{fig:varObsv}\,f.}}

In the case where the improper \textit{constant} $c_p$ model is applied within the model-based monitoring scheme, the estimates become biased as well.
This is true both for the individual estimates of $\aA$ and the overall thermal conductance $\kA$.
But in contrast to the model-free approach, the estimation error for $\kA$ is significantly smaller, and the Joint-EKF offers additional, useful information in terms of the so-called innovation \mbox{$\smallfcn{\vec{y}}{t} - \smallfcn{\vec{g}_\upsilon}{\vec{\hat{x}}_\upsilon,\vec{u},\vec{\theta},t}$}, which is the deviation between the measured and estimated outputs.
For the sake of clarity, we subtract the artificial measurement noise from the innovation, which leads to the ``noise canceled deviation" \mbox{$\smallfcn{\vec{\epsilon}}{\vec{x}, \vec{\hat{x}}_\upsilon} = \smallfcn{\vec{g}_r}{\vec{x},\vec{u},\vec{\theta},t} - \smallfcn{\vec{g}_\upsilon}{\vec{\hat{x}}_\upsilon,\vec{u},\vec{\theta},t}$}, i.e., the deviation between the true and estimated outputs, as depicted in Fig.\ \ref{fig:varObsv}\,d.
Every time the model-based estimation of $\kA$ is biased, the monitoring scheme indicates that deficit with a biased (not zero-mean) innovation (or deviation, here).
\hl{This is a result of the applied fluid model (constant $c_p$), which is improper for describing the reference counterflow process at that respective operating point.}

Obviously, the monitoring of $\kA$ is superior to a conventional (unfiltered) thermal rating\hl{\mbox{, cf.\ Fig.\ \ref{fig:varObsv}\,f.}}
This estimation is the result of the combined, estimated overall convection conductances $\smallfcn{\theta_1}{\hat{\p},\vec{\theta}_{hc}}$ and $\smallfcn{\theta_2}{\hat{\p},\vec{\theta}_{hc}}$, according to Eqs.\ (\ref{eq:kA_dyn}), (\ref{eq:aAh_Ansatz}), and (\ref{eq:aAc_Ansatz}).
Although their combination using Eq.\ (\ref{eq:kA_dyn}) shows satisfactory results, unfortunately, it does not apply to the convection conductances themselves\hl{, as can be seen in \mbox{Fig.\ \ref{fig:varObsv}\,e.}}
In numerous simulation studies, we noticed that the observability of these model parameters was highly sensitive to the exchanger's excitation.
\hl{For steady-state operation, they are not separately observable at all, but their combination $\kA$ is.
Hence, we state that the suggested monitoring scheme is incapable of offering reliable estimates of individual convection conductances. Thus, these estimates should be ignored for the monitoring task.}

\subsection{Monitoring with a Reduced-Information Setup}
\label{sec:reduced_information}

As a matter of fact, the assumed knowledge above concerning mass flows and intake temperatures as well as the existence of outlet temperature measurements, both on the hot and cold sides, provides the best possible conditions yielding the most reliable monitoring results.
Unfortunately, that premise often does not map the situation found commonly in industrial practice.
In this section, we aim to show the impact of unknown coolant flows $\md_c$ and canceled measurements of the coolant's outlet temperature $\T{c2}$ on the monitoring results.
Therefore, three variants of the Joint-EKF, adapted to a specific setup, are considered:
\begin{enumerate}
	\item[A)] assuming certain coolant flow knowledge and coolant outlet temperature measurements do exist,
	\item[B)] assuming uncertain coolant flow knowledge and coolant outlet temperature measurements do exist, and
	\item[C)] assuming uncertain coolant flow knowledge and coolant outlet temperature measurements do not exist.
\end{enumerate}
If $\md_c$ is uncertain, it is estimated by the Joint-EKF as well.
In contrast, if $\md_c$ is assumed to be known, the Joint-EKF uses a specified value or trajectory (here: $\md_c(t)=\SI{41}{\kilogram/\second} \ \forall t$) although the real trajectory might differ during a faulty situation.

Following the aforementioned Joint-EKF principle---so far, this is the joint estimation of $\vec{x}$ and $\p$, leading to design parameters $\mathbf{R_x}$ and $\mathbf{R_\upsilon}$---approaches B and C realize a joint estimation of $\vec{x}_\upsilon$ and $\md_c$, leading to design parameters $\mathbf{R_{x\upsilon}}$ and $R_{\md_c}$, where $\text{d}\md_c/\text{d}t \sim \smallfcn{\mathcal{N}}{0,R_{\md_c}}$.
Here, $\mathbf{R_x}$ and $\mathbf{R_\upsilon}$ are chosen as stated in (\ref{eq:Rp}); further, we set $R_{\md_c}=\SI{0.1}{\second}\cdot \left(\SI{1}{\kilogram/\second^2}\right)^2$.

\begin{figure}
	\begin{center}
		\fcolorbox{yellow}{yellow}{\includegraphics[scale=1]{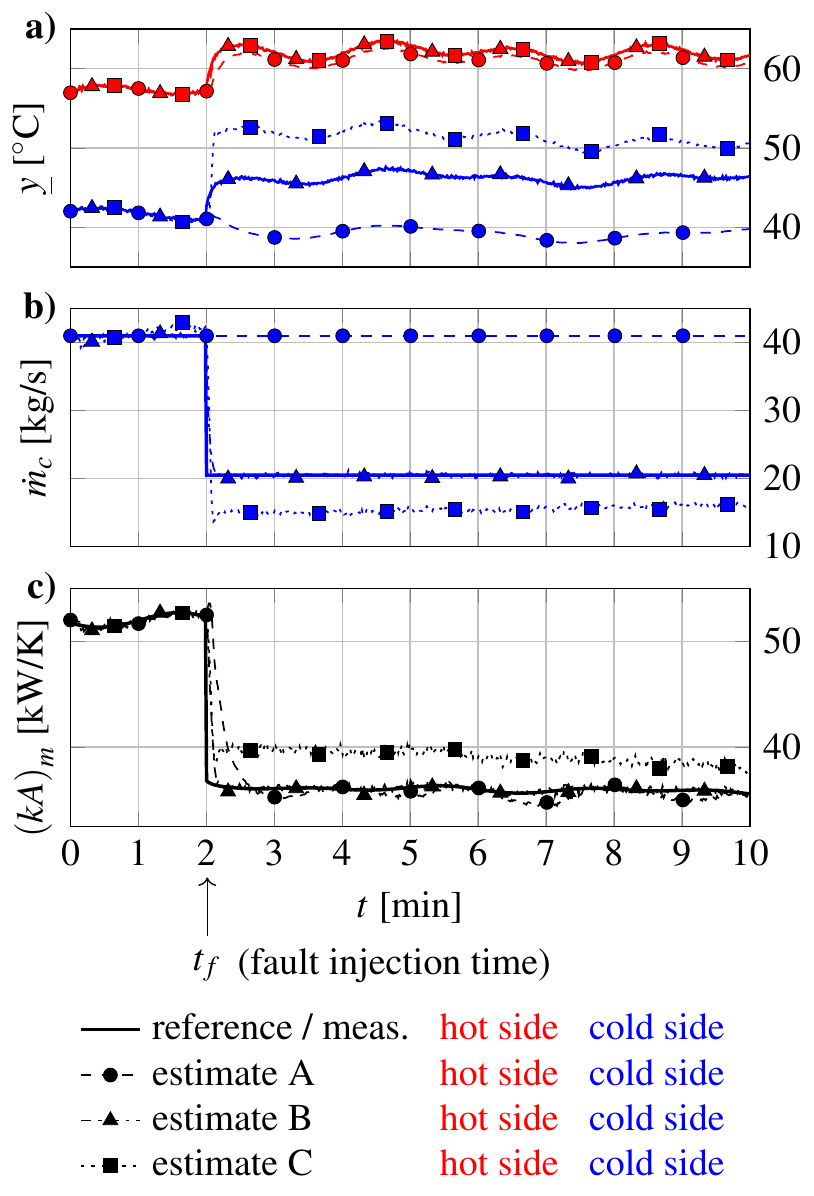}}
	\end{center}
	\caption[Monitoring with reduced information]{Monitoring with reduced information; \\ a) noisy measurements and respective model outputs; b) true and estimated coolant flows; and c) true and estimated overall thermal conductances;\\ estimate A) without coolant flow adaption, estimate B) with coolant flow adaption, and estimate C) without coolant outlet temperature feedback
	}
	\label{fig:unknown_mcStep}
\end{figure}

Again, the reference model is used to offer the ``true" \textit{reference} series, as depicted in Fig.\ \ref{fig:unknown_mcStep}.
Because we do not intend to discuss the impact of the chosen fluid model again, we use the same enthalpy calculation model within the reference and the monitoring model.\footnote{
	Carbon dioxide (process fluid): real gas model according to \cite{span1996new};  glycosol-water mixture (coolant): thermally perfect model according to \cite{glykosol}.
}
The simulated scenario is an abrupt coolant flow reduction at \hl{$t_f=\SI{2}{\minute}$} (magnitude: \SI{50}{\percent}), which is a critical failure in plant operations.
This time, some reasonable heat transfer correlations are used within the reference model (cf.\ Eq.\ (\ref{eq:Num})):\footnote{
	Mass flows and fluid properties are denoted with their respective SI unit.
	Fluid properties are calculated by a property database program \cite{lemmon2010nist}.}
 \begin{align}
\aAh &=  \SI{37}{\watt/\kelvin} \left(\frac{\md_h}{\SI{1}{\kilogram/\second}}\right)^{4/5} \left(\frac{c_{pm,h}}{\SI{1}{\joule/(\kilogram\ \kelvin)}}\right)^{1/3} \label{eq:aAh_Referenz}\\
&\hspace*{1cm} \cdot \left(\frac{\eta_{m,h}}{\SI{1}{\kilogram/(\meter\ \second)}}\right)^{-7/15}  \left(\frac{\lambda_{m,h}}{\SI{1}{\watt/(\meter\ \kelvin)}} \right)^{2/3}\, ,\nonumber
\\
\aAc &= \SI{2}{\watt/\kelvin}\left(\frac{\md_c}{\SI{1}{\kilogram/\second}}\right)^{4/5} \frac{c_{pm,c}}{\SI{1}{\joule/(\kilogram\ \kelvin)}} \label{eq:aAc_Referenz}\\
&\hspace*{1cm}\cdot \left(\frac{\eta_{m,c}}{\SI{1}{\kilogram/(\meter\ \second)}}\right)^{1/15}\, . \nonumber
\end{align}
Note that the mass flows and fluid properties are time-variant due to the time-variant inputs (not shown, with the exception of $\md_c$) and outputs of the reference model.
\hl{Due to the correlations stated above}, the abrupt coolant flow reduction causes a spontaneous breakdown of the \textit{reference} trend \hl{$\kA$} in Fig.\ \ref{fig:unknown_mcStep}\,c \hl{at $t=t_f$.}
In contrast, we use a simpler approach, exclusive of fluid properties, and with a different mass flow correlation within the monitoring setup, according to Eqs.\ (\ref{eq:aAh_Ansatz})--(\ref{eq:aAc_Ansatz}), with
 \begin{align}
\underline{\theta}_{hc} &= \begin{bmatrix} 0.6~ & 0~ & \SI{0}{\watt/\kelvin}~ & 0.6~ & 0~& \SI{0}{\watt/\kelvin} \end{bmatrix}^T \, .
\end{align}
Despite this biased approach, all estimates (A, B, C) performed well \hl{for $t<t_f$.}
None of the monitoring models receives information about the ``true" coolant flow trend, as this experiment should show the impact of uncertain inputs and reduced measurement information. 
As a matter of missing adaptability, estimate A assumes the preset coolant flow $\md_c=\SI{41}{\kilogram/\second}$ as a certain measure even after the event.
Thus, it is not able to adapt the correct physical cause (step of $\md_c$) in regard to the measured effect (behavior of $\vec{y}$). 
This applies especially to the cold side measurements, which are obviously unaccountable for with $\md_c=\SI{41}{\kilogram/\second}$, leading to a high deviation from estimate A.
Again, such model-based measures, i.e., the innovation in combination with the detected breakdown of $\kA$, could serve as meaningful indicators for \hl{fault detection.
Note that this work is related to the basic monitoring scheme that provides the parameter time series.
Thresholds or sophisticated fault detection approaches are not discussed herein.}

If an estimator is free to adapt the uncertain input $\md_c$, it is capable of staying on the true \textit{reference} tracks, as is the case for estimate B.
As a matter of course, it becomes worse if less measurements are considered. 
Estimate C provides proof of that.
Note that a \textit{steady calculation} on the basis of the heat transfer balance, like the mentioned rating in the previous section, would not be capable of compensating for two unknown or incorrect measures at once (here, $\md_c$ and $\T{c2}$) without a superior estimation technique.
To this point, one may interpret that setup C is proper for the monitoring purpose and that the Joint-EKF C only needs a little more time to converge to the correct values.
Unfortunately, we cannot guarantee this desired behavior for setup C.
Instead, we observed a higher sensitivity to the tuning matrices $\mathbf{R_x}$, $\mathbf{R_\upsilon}$, and $\mathbf{R_y}$ and a slow divergence rate (with a reasonable set of tuning matrices) concerning the estimates of $\kA$ and $\md_c$ for some simulation scenarios, especially if the system excitation was weak.
However, a drastic event, like the one shown, would have been detected by all approaches (A, B, C), at least in a qualitative manner.

\section{Conclusions}
\label{sec:conclusion}

For the purpose of monitoring a counterflow heat exchanger's thermal performance, we derived an appropriate model concerning the initially stated specifications.
Primarily, its ability to describe processes from steady to highly transient operating phases as well as the ease of deployment within industrial practice were the main demands.
The model considers the exchanger's dynamic behavior by only two ordinary differential equations.
Furthermore, the parametrization effort is very low.
More precisely, there are three to nine model parameters (depending on \textit{a priori} knowledge) and two arbitrary enthalpy calculation models, ($h_c(T,p)$ and $h_h(T,p)$), which are not restricted in their setup.
For the suggested monitoring scheme, one merely has to choose starting values for the model state and three additional tuning matrices, the effects of which on the monitoring behavior are easy to interpret.

To validate the model applied in the monitoring scheme, a reference model was derived.
It calculates model outputs on the basis of the current model states, parameters, and inputs in an accurate manner if the reference counterflow process is valid, as defined in Section \ref{sec:ReferenceProcess}.
The respective output equations are accompanied by high computational burden due to an integrated root determination.
This is why we derived approximate output equations that are solvable in one step.
Depending on the preset enthalpy calculation model and the specific root determination algorithm, the approximate model is substantially faster (about 50$\times$ for our setup), making it an eligible candidate for real-time applications.

Furthermore, the provision for an arbitrary enthalpy calculation model within the derived equations truly differs from conventional modeling approaches.
In this manner, the impact of an improper fluid model could be shown to result in biased observations.
As a result, the best-suited fluid model available should be used, which is mostly based on real gas equations.

The model-based estimation of unmeasurable system quantities is realized by a Joint-EKF approach, which is the joint estimation of the model states and parameters of the heat transfer correlation on the basis of the Extended Kalman Filter equations.
We were able to show that the chosen online estimation technique achieves admissible ratings of the overall thermal conductance for the fastest transients considered as well as for steady-state operating points.
Furthermore, a realistic situation of uncertain information (model inputs) and a reduced measurement setup were addressed.
We suggested an adapted estimation strategy for this setup.
As anticipated, the modified version was capable of compensating for one unknown input information smoothly, as this fact applies for a conventional rating based on the steady heat transfer balance as well.
An additional canceling of one outlet's temperature measurement reduces the reliability of the monitoring results since the estimator becomes sensitive to the tuning matrices and prone to (slow) divergence. 
This motivates a full instrumentation of heat exchangers.

In the present paper, we treated the heat exchanger as an isolated plant component with mostly known intake temperatures and flows.
The next step is to focus on a plant with integrated heat exchangers, where the input information of the presented monitoring algorithm arises from uncertain measurements and peripheral plant components.
The basic research presented here points out, again, the advantages of a model-based monitoring scheme.
The automated interpretation of the generated auxiliary measures (innovations and estimates), i.e., a fault-detection algorithm, will be a topic of our future research.

\section*{Acknowledgment}
This work was supported by MAN Energy Solutions SE and the Federal Ministry for Economic Affairs and Energy based on a decision by the German Bundestag as part of the ECOFLEX-Turbo project [grant number 03ET7091T].

\bibliography{bib__Gentsch_King__cooler_monitoring}

\appendix       
\renewcommand*\thefigure{\arabic{figure}}

\section{The uniqueness of the steady-state outlet temperatures}
\label{sec:appendixA}

Here, we argue why the root of 
\begin{numcases}{}
\fcn{\res{s1}}{\T{h2s}^*,\,\T{c2s}^*} := \fcn{\mathcal{H}_{c}}{\T{c2s}^*} + \fcn{\mathcal{H}_{h}}{\T{h2s}^*} \, , \nonumber \\
\fcn{\res{s2}}{\T{h2s}^*,\,\T{c2s}^*} := \fcn{\mathcal{H}_{c}}{\T{c2s}^*} &  \nonumber \\
\hspace*{2.2cm} - \fcn{\mathcal{Q}}{\T{h1}-\T{c2s}^*,\,\T{h2s}^*-\T{c1},\,\kA} &\nonumber 
\end{numcases}
is unique.
Let $\left(\T{h2s}^*,\, \T{c2s}^*\right) \in \left(\mathbf{A_1} \times \mathbf{B_1}\right)$ be a root of the set of roots $\left(\mathbf{A_1} \times \mathbf{B_1}\right)$ of $\res{s1}$.
One can easily find

$$\pd{\T{c2s}^*}{\T{h2s}^*} < 0\, , \quad \left(\T{h2s}^*,\, \T{c2s}^*\right) \in \left(\mathbf{A_1} \times \mathbf{B_1}\right)$$

due to physical reasonable positive heat capacities 

$$\pd{\fcn{h_{h/c}}{T,\,p}}{T}>0\ .$$

Thus, for a fixed $\T{c2s}^* \in \mathbf{B_1}$, there is a unique pair element $\T{h2s}^* \in \mathbf{A_1}$.
To determine the roots of $\res{s2}$ which are coexisting roots of $\res{s1}$ it is sufficient to vary one element of the given element pairs, e.g., $\T{h2s}^*$ within $\mathbf{A_1}$.
One can find the strict monotonicity

$$\pd{\fcn{\res{s2}}{\T{h2s}^*,\,\T{c2s}^*}}{\T{h2s^*}} < 0 \, , \quad \left(\T{h2s}^*,\, \T{c2s}^*\right) \in \left(\mathbf{A_1} \times \mathbf{B_1}\right)\, ,$$

which is sufficient proof of the uniqueness of the steady outlet temperatures $\left(\T{h2s},\, \T{c2s}\right)$.

%
%
\section{The uniqueness of the steady-state wall temperatures}
\label{sec:appendixB}

For the derivation of the low-order dynamic model equations, we postulate the uniqueness of a steady state, which is not a self-evident fact for nonlinear systems.
To at least emphasize the shown approach, we desired to sketch an elaborated proof of that prerequisite, which is very lengthly in its entirety.

In Section \ref{sec:steady_reference}, the steady-state wall temperatures $\left(\T{w1s},\, \T{w2s}\right)$ were introduced as the roots of
\begin{numcases}{}
\fcn{\res{s3}}{\T{w1s}^*,\,\T{w2s}^*} &\nonumber\\
\hspace*{1.0cm}:= \fcn{\mathcal{Q}}{\T{h1}-\T{c2s},\,\T{h2s}-\T{c1},\,\kA} &\nonumber \\
\hspace*{1.2cm} - \fcn{\mathcal{Q}}{\T{h1}-\T{w1s}^*,\,\T{h2s}-\T{w2s}^*,\,\aAh} \, ,&\nonumber \\
\fcn{\res{s4}}{\T{w1s}^*,\,\T{w2s}^*} &\nonumber \\
\hspace*{1.0cm}:= \fcn{\mathcal{Q}}{\T{h1}-\T{c2s},\,\T{h2s}-\T{c1},\,\kA} &\nonumber \\
\hspace*{1.2cm} - \fcn{\mathcal{Q}}{\T{w1s}^*-\T{c2s},\,\T{w2s}^*-\T{c1},\,\aAc} \, .&\nonumber
\end{numcases}
Note that $\T{h2s}$ and $\T{c2s}$ are predetermined and unique (see Appendix A).
In Fig.\ \ref{fig:SteadyState}, $f_3$ and $f_4$ are the graphs combining all of the roots of $\res{s3}$ and $\res{s4}$ in a separate manner:

 \begin{align*}
\fcn{\res{s3}}{\T{w1s}^*,\,\T{w2s}^*}&=0 &,\ \left(\T{w1s}^*,\, \T{w2s}^*\right) \in \left( \mathbf{A_3} \times \mathbf{B_3} \right) \, ,\\
\fcn{\res{s4}}{\T{w1s}^*,\,\T{w2s}^*}&=0 &,\ \left(\T{w1s}^*,\, \T{w2s}^*\right) \in \left( \mathbf{A_4} \times \mathbf{B_4} \right) \, .
\end{align*}
According to the qualitative situation in Fig.\ \ref{fig:SteadyState}, there is a unique shared root $\left(\T{w1s},\, \T{w2s}\right)$.
\begin{figure}
	\begin{center}
		\includegraphics{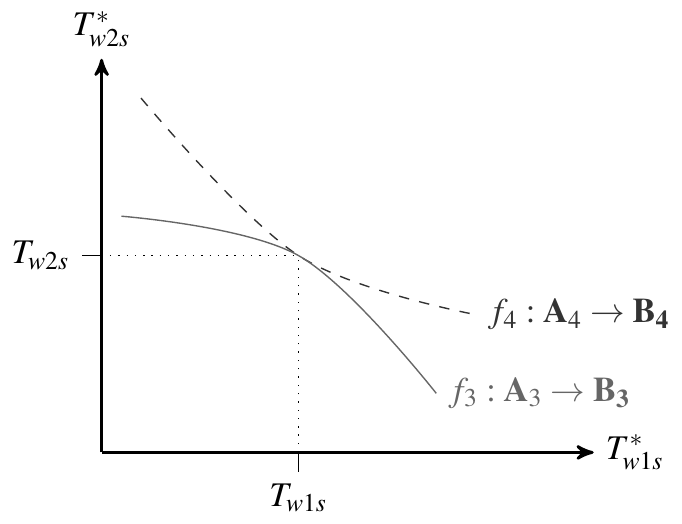}
	\end{center}
	\caption[The uniqueness of the steady-state]{The uniqueness of the steady-state}
	\label{fig:SteadyState}
\end{figure}
Obviously, the drawn graphs fulfill:
 \begin{align*}
\text{i)}&\qquad \left(\T{w1s},\, \T{w2s}\right) \in \left( \mathbf{A_3} \times \mathbf{B_3} \right) \, ,&& \\
\text{ii)}&\qquad \left(\T{w1s},\, \T{w2s}\right) \in \left( \mathbf{A_4} \times \mathbf{B_4} \right)  \, ,&& \\
\text{iii)}&\qquad \frac{d\, f_3}{d\, \T{w1s}^*} \bigg \vert_{\T{w1s}} = \frac{d\, f_4}{d\, \T{w1s}^*} \bigg \vert_{\T{w1s}}\, ,&& \\
\text{iv)}&\qquad  \frac{d^2\, f_3}{d\, \T{w1s}^{*\ 2}}\bigg \vert_{\T{w1s}^* \in \mathbf{A_3}} < 0 \, , \ \text{and} && \\
\text{v)}&\qquad \frac{d^2\, f_4}{d\, \T{w1s}^{*\ 2}}\bigg \vert_{\T{w1s}^* \in \mathbf{A_4}} > 0 \, . && 
\end{align*}
We can show that expressions iv--v are fulfilled in general and, further, with
 \begin{align*}
\T{w1s} &= \T{h1} + \frac{\aAc}{\aAh+\aAc}\cdot \left(\T{c2s}-\T{h1}\right)\, ,  \\
\T{w2s} &= \T{h2s}+ \frac{\aAc}{\aAh+\aAc}\cdot \left(\T{c1}-\T{h2s}\right)\, ,  
\end{align*}
the remaining expressions i--iii hold true.
Consequently, $\left(\T{w1s},\, \T{w2s}\right)$ is the unique shared root of $\res{s3}$ and $\res{s4}$.

\end{document}